\documentclass{article}

\pdfoutput=1

\usepackage{arxiv}
\usepackage[utf8]{inputenc} 
\usepackage{hyperref}       
\usepackage{url}            
\usepackage{graphicx}
\usepackage{amsmath,amssymb}
\usepackage{booktabs}
\usepackage[numbers,square,sort&compress]{natbib}
\usepackage{doi}
\usepackage{verbatimbox}

\title{Reversible Numeric Composite Key (RNCK)}

\author{ \href{https://orcid.org/0009-0003-8058-8413}{\includegraphics[scale=0.06]{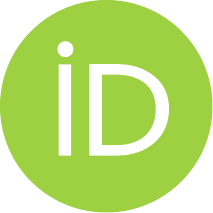}\hspace{1mm}Nicola Asuni}\thanks{\url{https://nicola.asuni.xyz}} \\
	\href{https://tecnick.com}{Tecnick.com LTD}\\
	Reading, UK\\
	\texttt{nicola.asuni@tecnick.com} \\
}

\hypersetup{
pdftitle={Reversible Numeric Composite Key (RNCK)},
pdfsubject={cs.DB, cs.DS},
pdfauthor={Nicola Asuni},
pdfkeywords={RNCK, Reversible Numeric Composite Key, composite key, data mining, database, decoding, encoding, hash, hashing, high-performance, hpc, software, surrogate key},
}

\begin{document}
\maketitle

\begin{abstract}
    In database design, composite keys uniquely identify records and prevent duplication. However, wide multi-column keys can increase index size, comparison work, and join costs. Surrogate keys can mitigate some of these costs, but they also require additional constraints and governance to preserve business-level uniqueness.

    This paper presents a \emph{Reversible Numeric Composite Key} (RNCK): a single non-negative integer that encodes multiple normalized attributes and can be decoded back to the original tuple under a fixed schema. RNCK is designed to combine the semantic fidelity of composite keys with the operational convenience of numeric keys.

    RNCK can improve storage footprint and key-comparison efficiency when attribute domains are bounded and stable. We formalize correctness and ordering properties, and specify operational semantics for partial-overflow mode. The approach has been used in production systems and is applicable to relational databases, static datasets, and key-value caching systems within the stated constraints.
\end{abstract}


\section{Introduction}
\label{sec:introduction}
In data modeling, or database design, a Composite Key is a unique identifier made up of two or more attributes (database table columns). For example, a patient record might be identified by the tuple (first\_name, last\_name, date\_of\_birth) in a constrained operational context where none of these attributes alone is guaranteed to be unique.

Composite keys generally help maintain data integrity and prevent data duplication in relational systems \citep{codd1970relational,silberschatz2019database}. Compared to surrogate keys, which are artificial identifiers that are not directly based on real-world attributes, composite keys can often be derived from existing natural attributes \citep{elmasri2016fundamentals}. However, when a composite key is referenced in multiple datasets (database tables) as a foreign key, it uses more memory or storage space, as multiple attributes (columns) are required instead of just one. This leads to a more complex schema. Queries can become more CPU-intensive, as every search and join requires comparing multiple attributes instead of a single key.

Surrogate keys can overcome some composite-key limitations at the expense of introducing other disadvantages. They do not encode business semantics and must be generated and maintained separately from the represented tuple, often together with additional uniqueness constraints on natural attributes \citep{kimball2013datawarehouse}.

The performance of queries, including memory consumption, is also affected by the type of each attribute.

Comparing fixed-width integer types is a highly optimized operation in modern computer architectures and database execution engines. In practice, throughput still depends on workload characteristics, memory hierarchy, and implementation details, so this paper focuses on formal properties, operational constraints, and applicability conditions.

In contrast with the simplicity and high-performance of numerical comparisons, string comparisons are a slow operation, especially for large strings. This is because strings are typically stored as arrays of characters and each character must be compared individually. This is assuming the best scenario where each string has already been normalized to a common canonical form. This is not necessarily common, more complex and expensive comparisons can be required. Alternatively, string comparisons can be performed more efficiently by using a hash table, but this can introduce some disadvantages like space complexity, collisions, load factor, and hash function performance \citep{cormen2022introduction}.

To combine the advantages of Composite Keys (CKs) and Surrogate Keys (SKs) while overcoming some of their limitations, a \emph{Reversible Numeric Composite Key} (RNCK) is presented here.

RNCK can be used only in certain cases, when the total number and maximum size of CK attributes is relatively small. RNCK encodes one or more attributes into a single non-negative integer, such that it is possible to directly and efficiently decode the original attributes while preserving some attribute sorting and searching properties. In these constrained settings, RNCK can improve query performance while reducing memory and storage requirements.

In addition to relational databases, RNCK can be a very effective index in large static datasets and key-value caching systems.

\section{Definition}
\label{sec:definition}
A Reversible Numeric Composite Key (RNCK) is a single non-negative integer that uniquely represents a Composite Key (CK) or a single non-numeric Natural Key.

Let the CK contain $n$ normalized attributes. For each attribute position $i \in \{1,\ldots,n\}$, let $\mathcal{A}_i$ be the finite set of admissible enumerated values. A composite key is the tuple
\[
\mathbf{a} = (a_1,\ldots,a_n) \in \mathcal{A}_1 \times \cdots \times \mathcal{A}_n =: \mathcal{A}.
\]
Let $\mathcal{K} = \{0,\ldots,2^w-1\}$ be the RNCK key space for a fixed machine word width $w$ in bits (typically $w=64$). The RNCK value is generated by an encoding function $E: \mathcal{A} \rightarrow \mathcal{K}$ and can be decoded back to the original attributes by a decoding function $D: \mathcal{K}_E \rightarrow \mathcal{A}$, where $\mathcal{K}_E := E(\mathcal{A})$.

For the bit-field construction used in the properties below, choose a bit allocation vector $\mathbf{b}=(b_1,\ldots,b_n)$ and define offsets $s_i := \sum_{j=i+1}^{n} b_j$. The canonical encoder is $E(\mathbf{a}) = \sum_{i=1}^{n} \nu_i(a_i)\,2^{s_i}$, where each $\nu_i : \mathcal{A}_i \to \{0,\ldots,2^{b_i}-1\}$ is a fixed injective enumeration map. The decoder is $D(k) = \bigl(\nu_1^{-1}(\pi_1(k)),\ldots,\nu_n^{-1}(\pi_n(k))\bigr)$, with $\pi_i(k) = \left\lfloor \frac{k}{2^{s_i}} \right\rfloor \bmod 2^{b_i}$.

\section{Normalization}
\label{sec:normalization}
Before encoding, a normalization step may be required to ensure a consistent and unambiguous representation of the CK attributes. For example, Unicode strings should be normalized to a canonical form.

Small in-memory lookup tables can be used by the encoding and decoding functions to efficiently enumerate limited attribute sets.

\subsection{Governance for Normalization and Enumeration}
To guarantee reproducibility across services and time, RNCK deployments should define and version a canonical preprocessing contract:
\begin{enumerate}
    \item \textbf{Canonical normalization pipeline.} Specify exact transformation order (e.g., trimming, Unicode normalization form, case handling, locale-independent formatting).
    \item \textbf{Null and sentinel policy.} Reserve explicit codes for null, unknown, and out-of-domain values.
    \item \textbf{Stable enumeration maps.} Treat each enumeration map $\nu_i$ as versioned data, not implicit application logic.
    \item \textbf{Migration rules.} When domains grow, introduce schema version bits or parallel encoders and backfill deterministically.
\end{enumerate}

\section{Format}
\label{sec:format}
In performance-focused applications, RNCK is typically a 64-bit unsigned integer (uint64), as this is natively supported by most current hardware platforms. Other data types compatible with binary operations can also be used, such as uint32 or uint128, if they are available.

An RNCK follows a binary pattern, where distinct sections of the binary number (groups of bits) represent different attributes or combinations of them. The binary sections are organized from the Most Significant Bit (MSB) to the Least Significant Bit (LSB) in the order of the sorting priority of each attribute. This allows sorting by RNCK to be equivalent to sorting by the attributes in order.

The number of bits required for each section follows Property~\ref{prop:bit-allocation}: for attribute domain size $|\mathcal{A}_i|$, the minimum valid allocation is $b_i=\lceil \log_2 |\mathcal{A}_i| \rceil$, and representability requires $\sum_i b_i \leq w$. For example, for an attribute with a maximum of 100 distinct values (including null), at least $\lceil \log_2(100) \rceil=7$ bits are needed because $2^7=128\ge100$.

Encoding and decoding complexity for the bit-field format is linear in the number of attributes with constant auxiliary space, as summarized in Property~\ref{prop:complexity}.

\section{Applicability and Limitations}
\label{sec:limitations}
RNCK can only be used instead of CK in certain cases, such as when the total number of attributes and the maximum size of each attribute is relatively small.

The use of RNCK is only possible if the underlying data type is large enough to contain the encoding of all CK attributes. Each binary section must be large enough to store all possible values of the corresponding attribute. This constraint is formalized in Property~\ref{prop:bit-allocation}. In some cases, additional flag bits may be required to indicate special cases.

\section{Partially Reversible Encoding}
\label{sec:nonreversible}
To overcome the RNCK capacity limitation, it is sometimes possible to adopt multiple encoding schemas that are indicated by bit flags.

For example, in VariantKey \citep{Asuni473744}, some input variants may exceed the REF+ALT binary section capacity. This is true for only about 0.4\% of the records in the reference dataset. In these rare cases, the least significant bit (LSB) is set to 1 and the remaining 30 bits are filled with a hash value that is used as a key for a relatively small lookup table. This alternate encoding is a good compromise because it is rarely used and still preserves some of the RNCK properties, such as the ability to sort and search the variants by chromosome (first binary section) and position (second binary section).
When the overflow key set is static, the auxiliary lookup table can be indexed with a minimal perfect hash function to avoid collisions in that table layer while keeping direct addressing \citep{botelho2007simple}.

For this partial mode, operational semantics should be explicit:
\begin{enumerate}
    \item \textbf{Collision handling.} If the overflow hash is not collision-free by construction, collisions must be resolved by deterministic secondary checks on stored normalized suffix attributes.
    \item \textbf{Lookup miss behavior.} A miss for a flagged key must be treated as a decoding failure (not silent fallback), with application-level handling defined in advance.
    \item \textbf{Integrity and versioning.} The overflow table and encoder version must be coupled; decoding requires the matching table snapshot.
    \item \textbf{Static vs evolving overflow sets.} Static sets may use MPHF-backed direct lookup, while evolving sets require updatable hash-index structures and migration procedures.
\end{enumerate}

This construction is summarized in Property~\ref{prop:partial-reversibility}.

\clearpage
\section{Properties}
\label{sec:properties}
\begin{enumerate}
    \item \label{prop:bit-allocation} \textbf{Bit allocation and width bound.} For each attribute $i$, the minimum number of bits that can encode all values in $\mathcal{A}_i$ is
    \[
    b_i = \left\lceil \log_2 |\mathcal{A}_i| \right\rceil.
    \]
    A $b_i$-bit field can encode at most $2^{b_i}$ values, so $2^{b_i}\ge |\mathcal{A}_i|$ is necessary. Choosing $b_i=\lceil\log_2|\mathcal{A}_i|\rceil$ is sufficient because an injective map into $\{0,\ldots,2^{b_i}-1\}$ then exists. RNCK is representable in $w$ bits exactly when $\sum_{i=1}^{n} b_i \leq w$.

    \item \label{prop:reversibility} \textbf{Reversibility, uniqueness, and injectivity.} If each $\nu_i$ is injective and $\sum_{i=1}^{n} b_i \leq w$, then for every admissible tuple $\mathbf{a}\in\mathcal{A}$,
    \[
    D(E(\mathbf{a})) = \mathbf{a},
    \]
    and for every encoded key $k\in\mathcal{K}_E$,
    \[
    E(D(k)) = k.
    \]
    The bit fields occupy disjoint ranges $[s_i,s_i+b_i-1]$, so projecting field $i$ from $E(\mathbf{a})$ recovers exactly $\nu_i(a_i)$. As a result, each admissible CK tuple has a unique RNCK code, each code in $\mathcal{K}_E$ decodes to a unique tuple, and distinct tuples cannot share the same RNCK value.

    \item \label{prop:ordering} \textbf{Ordering preservation.} If each $\nu_i$ preserves the intended attribute order, then numeric order on RNCK matches lexicographic order on CK tuples:
    \[
    \mathbf{a} <_{\mathrm{lex}} \mathbf{a}' \iff E(\mathbf{a}) < E(\mathbf{a}').
    \]
    If $t$ is the first index where $a_t\neq a_t'$, then all higher-priority field contributions cancel. Field $t$, weighted by $2^{s_t}$, determines the sign of the difference because the total contribution of lower-priority fields is strictly smaller than $2^{s_t}$.

    \item \label{prop:hex-ordering} \textbf{Hexadecimal ordering.} For fixed-width hexadecimal rendering, zero-padding to $w/4$ digits preserves numeric order. Combined with Property~\ref{prop:ordering}, alphabetical order on the hexadecimal representation matches CK lexicographic order.

    \item \label{prop:complexity} \textbf{Encoding and decoding cost.} Computing either $E(\mathbf{a})$ or $D(k)$ takes $O(n)$ time and $O(1)$ auxiliary space for fixed machine word width. Encoding performs a constant number of primitive operations per field, while decoding uses one shift and one mask per field plus constant-time reverse lookup over the finite domain model.

    \item \label{prop:comparison-cost} \textbf{Comparison and storage cost.} Comparing two RNCK values is $O(1)$ time because it is a single machine-word comparison. Comparing CK tuples is $O(m)$ in the worst case, where $m$ is the number of compared attribute bytes examined before the first mismatch. RNCK storage is $O(1)$ machine words, while CK storage is $\Theta\!\left(\sum_i |a_i|\right)$ bytes before compression.

    \item \label{prop:joins} \textbf{Join semantics preservation.} Replacing CK tuples with RNCK values preserves the result of inner, left, right, and full joins when both datasets use the same normalization and encoding schema. Join semantics depend only on key equality, and Property~\ref{prop:reversibility} makes CK equality equivalent to RNCK equality under a shared encoder.

    \item \label{prop:hash-table} \textbf{Hash-table compatibility.} RNCK values are valid hash-table keys with expected $O(1)$ lookup under standard randomizing hash assumptions. A fixed-width integer domain can be handled by standard integer hash functions, and entropy present in the concatenated fields remains available to the hash mixer.

    \item \label{prop:partial-reversibility} \textbf{Partial reversibility with prefix preservation.} Let
    \[
    E_{\mathrm{partial}}(\mathbf{a})=(E_{\mathrm{prefix}}(a_1,\ldots,a_t),\; h(a_{t+1},\ldots,a_n),\; f),
    \]
    where $f$ is a mode flag and $h$ is a hash used with an auxiliary lookup table. Ordering by the encoded prefix is preserved exactly as in Property~\ref{prop:ordering}, while full reversibility holds only for rows resolved through the lookup table. Prefix bits remain unchanged from the fully reversible scheme, whereas the hashed suffix is generally not injective.

    \item \textbf{Columnar compatibility.} RNCK is compatible with columnar data formats such as Apache Arrow \citep{ApacheArrow} and Apache Parquet \citep{ApacheParquet}, where fixed-width numeric keys support efficient binary search and vectorized filtering.
\end{enumerate}

\section{Related Work}
\label{sec:relatedwork}
RNCK is adjacent to several established techniques but targets a distinct combination of properties.

Bit packing and field concatenation are common in systems programming and columnar execution paths, where compact fixed-width representations improve memory locality and scan performance \citep{abadi2006integrating}. RNCK adopts this representation style while emphasizing formal reversibility for bounded domains.

Dictionary encoding and surrogate-key strategies are widely used to reduce storage and accelerate joins in warehousing and analytics systems \citep{kimball2013datawarehouse}. RNCK differs by requiring deterministic decode to the original normalized tuple (or explicit partial mode when overflow is used), rather than only preserving identity through a separate lookup.

Partially reversible constructions with hashed suffixes relate to minimal perfect hashing and compact lookup structures \citep{botelho2007simple}. RNCK contributes an ordering-preserving prefix layout and an explicit operational contract for overflow semantics.

\section{Examples}
\label{sec:examples}
Practical implementations of Reversible Numeric Composite Key (RNCK) in multiple programming languages have been successfully used in production systems for some time.

\subsection{VariantKey}
VariantKey is an RNCK for Human Genetic Variants \citep{Asuni473744}.

A reference implementation of VariantKey in multiple programming languages is available in the project repository \citep{GitHubVariantKey}.

In formal notation, VariantKey instantiates $n=3$ with tuple $\mathbf{a}=(\text{CHROM},\text{POS},\text{REF+ALT})$ and bit allocation $\mathbf{b}=(5,28,31)$, satisfying Property~\ref{prop:bit-allocation} and preserving lexicographic order by chromosome then position (Property~\ref{prop:ordering}).

\noindent\begin{minipage}{\linewidth}
\emph{The VariantKey is composed of 3 sections arranged in 64 bit:}
\begin{verbnobox}[\small]
         0   4 5                             32 33                              63
         |   | |                              | |                                |
         01234 567 89012345 67890123 45678901 2 3456789 01234567 89012345 67890123
5 bit CHROM >| |<         28 bit POS         >| |<        31 bit REF+ALT        >|
\end{verbnobox}
\end{minipage}

\noindent\begin{minipage}{\linewidth}
\emph{Encoding example:}
\begin{verbnobox}[\small]
                   | CHROM | POS                          | REF | ALT                           |
-------------------+-------+------------------------------+-----+-------------------------------+
       Raw variant | chr19 | 29238770                     | TC  | TG                            |
Normalized variant | 19    | 29238771                     | C   | G                             |
-------------------+-------+------------------------------+-----+-------------------------------+
    VariantKey bin | 10011 | 0001101111100010010111110011 | 0001 0001 01 10 0000000000000000000 |
-------------------+-------+------------------------------+-------------------------------------+
    VariantKey hex | 98DF12F988B00000                                                           |
    VariantKey dec | 11015544076520914944                                                       |
-------------------+----------------------------------------------------------------------------+
\end{verbnobox}
\end{minipage}

\subsection{NumKey}
NumKey is an RNCK for Short Codes or E.164 LVN.

A reference implementation of NumKey in multiple programming languages is available in the project repository \citep{GitHubNumKey}.

In formal notation, NumKey instantiates $n=3$ with tuple $\mathbf{a}=(\text{COUNTRY},\text{NUMBER},\text{LENGTH})$ and bit allocation $\mathbf{b}=(10,50,4)$, so $\sum b_i=64$ and full reversibility follows from Property~\ref{prop:reversibility}.

\noindent\begin{minipage}{\linewidth}
\emph{The NumKey is composed of 3 sections arranged in 64 bit:}
\begin{verbnobox}[\small]
    0         9 10                                                    59 60 63
    |         | |                                                      | |  |
    01234567 89 012345 67890123 45678901 23456789 01234567 89012345 6789 0123
10 bit COUNTRY >| |<                     50 bit NUMBER                  >| |< 4 bit LENGTH
\end{verbnobox}
\end{minipage}

\noindent\begin{minipage}{\linewidth}
\emph{Encoding example:}
\begin{verbnobox}[\small]
               | COUNTRY    | NUMBER                                          | NUM |
               | [ISO 3166] | [E.164]                                         | LEN |
---------------+------+-----+-------------------------------------------------+-----+
       Number  |   I     T  | 123456                                          |  6  |
---------------+---- -+-----+-------------------------------------------------+-----+
    NumKey bin | 10011 01000 0000000000000000000000000000000011110001001000000 0110 |
---------------+--------------------------------------------------------------------+
    NumKey hex | 4D000000001E2406                                                   |
    NumKey dec | 5548434740922426374                                                |
---------------+---+----------------------------------------------------------------+
\end{verbnobox}
\end{minipage}

\section{Conclusions}
Reversible Numeric Composite Key (RNCK) is a key-construction method that combines properties of composite and numeric surrogate representations while preserving decode capability under bounded-domain assumptions. RNCK preserves useful sort and search properties when enumeration maps are stable and schema constraints are satisfied.

RNCK can only be used in certain cases, such as when the total number and maximum size of key attributes is relatively small. In these cases, RNCK can be applied effectively to relational databases, static datasets, and key-value caching systems.

Adopting RNCK can reduce memory and storage requirements and can improve query performance in suitable workloads. The presented formal properties and operational constraints are intended to support rigorous validation in future empirical studies.

\bibliographystyle{unsrtnat}
\bibliography{rnck} 

\end{document}